# TOWARDS A RESILIENT INFORMATION SYSTEM FOR AGRICULTURE EXTENSION INFORMATION SERVICE: AN EXPLORATORY STUDY

Muluneh Atinaf, Addis Ababa University, Ethiopia, mulunehatinaf@yahoo.com

Alemayehu Molla, RMIT University, Australia, alemayehu.molla@rmit.edu.au

Salehu Anteneh, Addis Ababa University, Ethiopia, salehu.anteneh@aau.edu.et

**Abstract:** Although digital technologies are contributing to human development, several information systems (IS) interventions for development especially in developing countries do not perform as expected nor deliver anticipated outcomes at scale. This raises questions about how to develop and enhance resilient IS for development, an area that requires more research attention. A sound and systematic understanding of the mechanisms local communities apply to maintain resilience and the key transformation areas for a resilient IS development will help to improve the situation. This study addresses how stakeholders can ensure resilient information provision within the Agricultural Extension Information Service (AEIS) and identifies the challenges in designing resilient IS. Conceptually, the study draws from the IS resilience framework. Empirically, it draws from interview data collected from the AEIS provision practice in Ethiopia. The findings show the robustness, self-organization, learning, redundancy, rapidity, scale, diversity and equality mechanisms, the challenges and the key transformations required to advance the resilience of IS for AEIS. The study contributes to the conversation on the application of the IS resilience framework in analyzing local information provision practices as well as to practice highlighting the key transformation areas to improve the effectiveness and impact of AEIS.

## 1. INTRODUCTION

The contribution of information and communication technology for development (ICT4D) is widely acknowledged in the literature, although its role in achieving sustained impact remains debatable. Some ICT4D projects fail (Heeks, 2002) and/or lack critical resilience properties (Heeks & Ospina, 2018). To address this, several approaches have been tried. For instance, participatory design strategies that define users' right to be involved in the process of design and that give some power to users (Carroll & Rosson, 2007) are proved to be firm-centric and not ideal for IS solutions in the ICT4D domain (Walsham, 2010). Others focused on analyzing how ICT can foster "development" (Sahay et al., 2017); improving inclusion and participation of marginalized subjects (Masiero, 2018); and context-specific ICT4D theorizing (Qureshi, 2015; Avgerou, 2017; Andoh-baidoo, 2017). While these efforts have contributed to the advancement of the ICT4D sub-field, research on resilience within the ICT4D field is limited (Heeks and Ospina, 2018). In the same token our understanding of the mechanisms that would help to figure out building resilient IS for AEIS is limited. This is because ICT4D is embedded in a multi-faceted socio-technical reality that involves several stakeholder interactions across structures (Atinaf et al., 2020). This complexity requires unraveling resilience maintaining mechanisms, challenges and transformations, an area that needs further investigation particularly in the agricultural context, which is the main stay of economy and employment for most developing economies.

Specifically, the contributions of local practices towards AEIS resilience, the key transformations required as well as the challenges towards problematizing the development of resilient information systems for AEIS has yet to be investigated. This study addresses this need guided by the following research question: How do stakeholders in an AEIS system ensure resilient information provision and what are the challenges to design resilient IS for AEIS? The objective is to identify how stakeholders ensure resilient information provision within the agricultural extension service by overcoming the challenges through transformative actions to develop resilient IS for AEIS. Such an understanding is important both for identifying relevant systems affordances and deciding the scope and nature of future interventions (Lewis, 1992).

The AEIS interest for this research is in Ethiopia, where agriculture is the backbone of the country's' economy and the source of livelihood. It constitutes more than half of the country's GDP; accounts for over 80% of the labor force; and is a major source of export earnings (Atinaf et al 2020). The AEIS in Ethiopia is designed to deliver a wide variety of agriculture related information services and is characterized by a complex service ecosystem with a range of stakeholders that differ in education, participation, and capacity (Barau & Afrad, 2017; Atinaf et al., 2020). The Ethiopian AEIS eco-system therefore provides an ideal setting to explore the mechanisms local agents and practitioners apply to enable the resilience of information services provision within a multi-stakeholder environment. It also offers the opportunity to reflect on how the circumstances of a particular context influences and is influenced by the attributes of resilience (Heeks & Ospina, 2018; Davison and Andrade, 2018).

The rest of the paper is organized as follows: the next section provides a background to the domain of IS resilience. This is followed by description of the research methodology employed in our empirical study. Next, we present the findings and results from analysis of the empirical data. We then discuss the results and findings based on the resilience attributes and markers. The final section presents the implications of our study to a resilient AEIS system.

## 2. BACKGROUND LITERATURE

Both supply and demand considerations that balance institutional and technical elements are important to enhance the sustainability of ICT4D initiatives (Pade-Khene & Lannon, 2017). Nevertheless, the sustainability of ICT4D interventions are challenged due to supply driven perspectives that focus on augmenting the supply side resources and that develop IS that reflect the values and views of suppliers (Sahay & Mukherjee, 2017). Supply driven perspectives introduce complex dependencies on resources, ignore the agency of users, the role of institutions, and many other demand-side conditions (Sahay & Mukherjee, 2017) and do not sufficiently focus on the resilient attributes of information systems (Heeks & Ospina, 2018) that are critical to offer sustainable services. This indicates that the concept of resilience has not received much attention in the ICT4D domain, where the focus has been more on one-way or two-way communications, digitizing existing flows and processes than reengineering and supporting multi-stakeholder networks (Heeks & Ospina, 2018). This approach is against the trends in agricultural research and development process which emphasize the importance of multi-stakeholder engagement and participatory approaches (Adekunle & Fatunbi, 2012).

The concept of resilience in IS is multi-dimensional implying a perfect reliability (that is free of failure) is impossible (Sakurai & Kokuryo, 2014). However, the situation can be improved by focusing on the need for resilient IS (Sakurai & Kokuryo, 2014) that can cope with disturbances at different stages (Müller et al., 2013). Thus, resilience should not be viewed as a static phenomenon but something to be improved and developed through time. IS research has also seen resilience from three important perspectives. The first views it from a technical perspective with a focus either on the post disaster property, such as recoverability, of a system (Wang et al., 2010; Zhan &

Lin, 2010; Sakurai & Kokuryo, 2014) or on "what is left" after the damage, that is, the redundancy of a system (Ash & Newth, 2007; Sakurai & Kokuryo, 2014). This perspective views resilience as systems' ability to regain important capabilities for a smooth recovery of operations post disaster (Sakurai and Kokuryo, 2014). The second looks at resilience as the ability of a system to cope with the before, during, and after disturbances (Müller et al., 2013). The third views resilience with a wider scope giving attention to ICT4D and the current development agendas to address inequality within a social system (Heeks and Ospina 2018).

This paper follows conceptualization and framework of resilience for the ICT4D domain as the "ability of a system to withstand, recover from, and adapt to short-term shocks and longer-term change" and one that avoids continuing systemic inequalities (Heeks and Ospina, 2018, p6). Avoiding systemic inequalities is particularly relevant for poor farmers' access to agriculture extension and information services. Heeks and Ospina's (2018) resilience framework is composed of foundational and enabling attributes. The foundational attributes consist of robustness, self-organization, and learning and the enabling attributes are composed of redundancy, rapidity, scale, diversity and flexibility, and equity attributes. Heeks and Ospina (2018) also identify markers that characterize each of these attributes (Table 1).

| Resilience attributes | | Description | Markers |
|---|---|---|---|
| Foundational Attributes | Robustness | Ability of the system to maintain its characteristics and performance in the face of contextual shocks and fluctuations | Physical preparedness; Institutional capacity; Loose functional coupling |
| | Self-organization | Ability of the system to independently rearrange its functions and processes in the face of an external disturbance | Collaboration and consensus-building; social networks and trust; local leadership |
| | Learning | The systems' capacity to generate feedback to gain/ create knowledge, strengthen skills and capacities necessary to experiment and innovate | Capacity building; new and traditional knowledge; reflective thinking |
| Enabling attributes | Redundancy | The extent to which components within a system are substitutable | Resource sparseness; resource substitutability; functional overlaps |
| | Rapidity | Accessing and mobilizing assets quickly to achieve goals in an efficient manner to ensure the system's ability to respond to external stressors timely | Rapid issue detection; rapid issue assessment; rapid issue response (resource mobilization) |
| | Scale | The breadth of assets and structures a system can access to effectively overcome/bounce back from/adapt to the effects of disturbances | Scale of resource access; multilevel networks; intra-level networks |
| | Diversity and flexibility | Availability of a variety of assets knowledge, institutions, and institutional functions that enable a range of response options to external stressors, both short- and long-term | Variety of courses of action; adaptable decision-making; innovation mechanism |
| | Equality | The extent to which the system affords equal access to rights, resources and opportunities to its members | Equality of distribution of assets; inclusiveness and participation; openness and accountability |

**Table 1: Framework of Resilience (adapted from Heeks & Ospina, 2018)**

The AEIS is a multi-stakeholder information service. It requires identifying key players (Haythornthwaite, 1996) who would champion the ICT initiatives (Renken & Heeks, 2018) to justify the continued existence of information (Haythornthwaite, 1996). Hence, based on Heeks and Ospina (2018) framework, this research will explore the current mechanisms of information service provision in the Ethiopian AEIS, the prevalent challenges and issues and the key transformations to improve the resilience of IS for AEIS.

# 3.   RESEARCH METHOD

The study is based on a qualitative case study as such an approach gives an advantage to focus on the real-life experiences and contexts of the phenomenon (Yin, 2009; Eisenhardt & Graebner, 2007) in environments where users' context of action is critically important (Benbasat et al., 1987). Data are collected as part of a wider project, partially presented in Atinaf et al (2020), from the stakeholders involved in the AEIS system.

As reported in Atinaf et al (2020, p 5) "qualitative data were collected through interviewing a range of stakeholders at one AEIS system locale from the "Amhara" Regional State of Ethiopia. Ethiopia is [administratively] divided into different administrative structures called regions, which are further divided into administrative zones. A zone has a number of "Woredas" in it. The "Woreda" administrative unit is composed of "Kebeles", the lowest units in the administrative. The [AEIS] is led by the Federal Democratic Republic of Ethiopian (FDRE) Ministry of Agriculture and is typically composed of the Regional State Agriculture Bureau, Zonal Agriculture Office, "Woreda" Agricultural Extension Office, Development Agents (DAs), Farmers' Associations, and Farmers." Other organizations are involved in supporting the services, these include, "research centers, local cooperative offices, seed producing organizations, and Unions of Farmers Associations. The [AEIS] involves the above stakeholders operating at different levels where farmers are the primary beneficiaries of the services. Data collection took place in early 2017 by using theoretical sampling approach complemented with the snowballing technique to identify potential interviewees. The interviews started from the Director and Agricultural Extension Department Head of the "Amhara" Region Agricultural Research Institute to discuss the overall AEIS in the region.

In total, 20 individual and group interviews with 29 participants covering the Heads of local agriculture extension and woreda cooperative offices, development agents, farmers, and other key informants from farmers association, research centers and head of Information Technology unit at the Agricultural transformation Agency (ATA) are conducted. See Atinaf et al (2020, p 5) for profile of interviewees.

Data are first transcribed and then translated from Amharic language to English language. The translated data is checked against the transcription by a language expert to ensure meaning was not lost during translation. Data analysis starts with thematic coding technique (Saldaña, 2009) following the foundational and enabling attributes of resilience (Heeks & Ospina, 2018)

# 4.   RESULTS

The findings are presented with three primary goals: to identify the mechanisms the AEIS community applies to ensure foundational and enabling resilience attributes, the challenges encountered, and the transformations needed to overcome these challenges. This is followed by a discussion of the resilience attributes and resilience markers.

### 4.1. Foundational Resilience Attributes

### 4.1.1. Robustness

Erratic weather conditions and pests are some of the shocks the AEIS stakeholders face. These shocks which are not planned a priori in the AEIS system are addressed by applying community information sharing practices. For instance, informants from Development Agents, the Research Center, and Research Institute mentioned that, even if the ATA pushes information from the center to the stakeholders, farmers have their own mechanisms of producing pesticides and have convincing arguments regarding their agricultural practices such as sowing seeds and farmland plowing to fight crop disease. Moreover, the farmers use their social structure and local knowledge to manage unplanned disturbances such as shortage of expert support. For instance, an expert from a Woreda office indicated

> *… it will take us two days to give support for smallholder farmers in a single rural kebele when most of them will be engaged in similar and seasonal activity. We will use farmers' development group leaders as experts to give support to others and communicate the information to be applied by the smallholder farmers.*

### 4.1.2. Self-organization

The AEIS system is organized as a self-helping community to respond to the problems that its members face. One of these is its structure and the defined roles that members have within this structure. The stakeholders include the farmers', development agents, farmers' associations, subject matter specialists, researchers, and farmers' development groups. The roles of these stakeholders are presented in Table 2. The stakeholders participate in exchanging contextually relevant information tailored to individual fields, taking the divergent concerns of the stakeholders into context through discussions and supervision, continuous interactions, and via their growing actor networks. The stakeholders apply these defined roles to re-arrange themselves as self-support communities in order to put the functions of the agricultural extension information provision functions in place.

| Key Actor | Role in Agricultural Extension Information Service |
|---|---|
| Farmers | Use information; identify and communicate experiences, concerns, challenges, and issues and problems to adopt new technologies |
| Development agents | Information brokering and transferring development work programs to farmers, communicating concerns, challenges, issues, and technology priorities of farmers to SMS and researchers, provide feedback on technology to SMS and researchers |
| Farmers' association (FA) | Announcing market price information and marketing schedule to farmers; communicate availability of agricultural input and other machinery technologies to farmers |
| Subject matter specialists | Providing information to farmers and DAs; communicate solution information to problems and issues, communicate development work programs to DAs |
| Researchers | Identifying farmers' technology needs and technology intake and creating awareness on the technologies, information to DAs and SMS via trainings |
| Development groups (DG) | Develop and communicate concerns, challenges, and issues of smallholder farmers via group leaders; assist and support farmers by providing information in the absence of DAs; disseminate input and other agricultural technology needs of farmers to DAs |

**Table 2. Actor in the agriculture extension service**

### 4.1.3. Learning

The AEIS community has its own ways of capacity building mechanisms such as the use of the social structure of the farmers to enable learning from each other. There is horizontal and vertical information exchange across the structures that involve individuals, institutions, and groups/collectives to facilitate the learning via the community's local languages. This is used as a strategy to empower members of the stakeholders in the AEIS system with skills and knowledge of agricultural practices. The mechanisms applied include direct observation, discussions, trainings, meetings, joint activities, supervision. A development agent in an interview indicated,

> *… we will attend meetings with the smallholder farmers whenever they assembled together for their own purpose and make discussions on issues related to their farming practices.*

Stakeholders apply farmers' field observations, group discussions to devise solutions for their own problems, organize trainings on selected issues, and perform activities in groups, and supervision of smallholder farmers and other stakeholders along the social and institutional hierarchy.

## 4.2. Enabling Resilient Attributes

### 4.2.1. Redundancy

Substitutability of access to agricultural extension information is ensured via applying various information delivery mechanisms. These mechanisms offer advantage of redundant access to agricultural extension information which at the same time provides inclusive access to the needed information by members. The information exchange mechanisms applied include word-of-mouth public announcements, service encounters, publications on paper and web portals, supervision, training, broadcasting over radio and television, discussions, and joint activities. The use of mobile phones and the ATAs' Interactive Voice Response (IVR) system is a prime example to the substitutability of the AEIS. For instance, an expert at the Woreda office indicated the following:

> *Many farmers have mobile phones with them or household members, and we use mobile phone calls to communicate urgent information with stakeholders.*

These information exchange mechanisms enable redundancy of information provision functions with different forms of presentation, access from different sources, and through different devices to improve availability or reliability of AEIS. The stakeholders in the AEIS system make scare resources abundant via these mechanisms to address problem of dependency on single source.

### 4.2.2. Rapidity

The speed of information access, assessment and mobilization of information is routed through mobile phone calls as well as the social and institutional structures. Stakeholders apply mobile phone calls to meet the urgent needs of information. An interviewee from the development Agent indicated that

> '*farmers usually call over our mobile phones when they need information urgently.*"

However, it doesn't mean that every smallholder owns a mobile phone and has access to experts in the AEIS, a challenge in fulfilling rapidity. The social structures within the farmers' villages are instrumental to these communities in routing agricultural extension information rapidly to members. The Development Groups in the rural Kebeles facilitate the flow of information within the AEIS system so that everyone has access to contextual information needed. This structure is organized based on gender and age into the men development groups, women development groups, and youth development groups. Smallholder farmers use this semi-social structure as a platform to circulate information rapidly in addition to using it as a learning mechanism in the AEIS system. Moreover, the farmers associations formed by farmer members are used to disseminate market related information to members. Board members and farmers who take leadership positions in the farmers associations and the rural Kebeles take the prime responsibilities in communicating such information.

### 4.2.3. Scale

Agricultural extension information is contributed and used by different stakeholders. The individual farmers, farmers' development groups with its multi-level structure, the development agents, subject matter specialists at a higher administrative level, the researchers from research institutes, and the farmers associations are involved in the information exchanging process. Members within this social and institutional networks interact vertically (top-down and bottom-up) across structures and horizontally within a specified structure. The social and institutional structures can be easily scaled up and grow to form networked actors. This network is available to respond to stakeholder's queries of help and support related to AEIS both to the members as well as to others across the structure. The stakeholders are connected each other to enable the network to grow or scale and functions at any scale within the administrative structure.

#### 4.2.4. Diversity and Flexibility

Diversity and flexibility of agricultural extension information exchange is enabled through the application of different mechanisms. For instance, the ATAs' IVR-based system which delivers agricultural extension information via mobile devices, the social structures within the stakeholders which acts as a self-help group, and the use of other forms of extension information communication are prime examples. Other agricultural extension information provision mechanisms that mark flexibility and diversity include word-of-mouth public announcements, mobile phone call, service encounters, publication, supervision, training, broadcast media, discussions, and joint activities. For instance, a farmer indicated that:

> *… when I hear important information, I will announce it to the villagers. Board members of the farmers associations do the same thing to circulate information that should be addressed to other farmers.*

However all forms of these diversity and flexibility markers is not suitable to all stakeholders. A project director at a university working to enhance knowledge of DAs asserted the situation as follows:

> *Information and knowledge exchange platforms, such as the Ethiopian agricultural web portal, is highly scientific operating at a national level in an environment where agricultural practice is affected by different ecological and social factors.*

#### 4.2.5. Equality

The extent of equal access to AEIS cannot be described based on the empirical data as there is a precise marker for it. However, there are markers in the AEIS system that shows equal access to information. Agricultural extension information is a needs-based service and offered based on stakeholders' desires. The organization of the development groups at different levels, constructing service encounters at Kebele offices, and a volunteer-based farmer associations are markers for equal access to AEIS. Though it is a volunteer-based public service, there are challenging conditions facing smallholder farmers in accessing expert personnel when there is a need to.

### 4.3. Challenges and Issues in the AEIS System

The findings also show some persistent issues that challenge resilience of the AEIS including the current IS for AEIS. These are related to existing agricultural IS and IS development methodologies, access to basic AEIS, digital literacy, and access to market information.

#### 4.3.1. Information Systems Development Methodologies

Challenges and issues in the current computer-based IS being used for AEIS are categorized into two. The first is related to lack of tools for guiding the design and development of the information systems. Supporting this respondent at the ATA told the following:

> *In our country, there is a culture of designing IS based on what comes to our mind all of a sudden; it is the same in this ATA project. The project doesn't follow software development/software engineering principles, lacks documentation, and there are problems related to integration and scale up.*

The second is reflected in the existing IS and the extent to which they incorporate information from stakeholders, facilitate automatic and effective information transfer, and are sufficiently stocked with relevant agriculture extension information. Information received by the smallholder farmers are hardly sourced from farmers themselves. For instance, although farmers have developed their own pesticides and arguably effective farming practices, existing IS designs are limited to include such information into current agricultural extension practices.

Poor information transfer is also manifested in the IS content stored in the ATAs' system as it relates to meeting the stakeholders needs. Supporting this, a developer at ATA reported that, *'content stored in the IS is updated every six months and provides too general recommendations'*.

Moreover, the system is criticized for lacking features to support the stakeholders' roles in their day-to-day activities. The challenge of information transfer is also associated to transfer of insufficient information packages and recommendations related to farming technology. An informant from the Research Institute stated this as:

*For a specific technology, millet, applying the standard package of the technology in the research center produced 40 quintals per hectare. The same amount of produce was gained in the farm trial phase too. When the technology is distributed to the farmers, the total produce from the same size of land was reduced to 10 quintals. The reason was inappropriate application of the technology package.*

### 4.3.2. Lack of/Limited Access to Basic Agricultural Extension Services

Lack of/ or having limited access to basic AEIS is another challenge affecting the effort made to offer effective public AEIS. The proportion of service providing personnel and institutions to that of smallholder farmer's population, geographic distance, and commitment of those personnel and organizations remain the main limiting factors creating a divide. The empirical data also revealed that the gap gets worse during high agricultural activity seasons (for instance during sowing natural resources protection). Moreover, our data shows, those who are geographically near to experts or institutions have better access than those located at a distant. A farmer noted the extent of geographical barrier as

*"the problem is serious for female DAs who cannot walk for long distance and cross dense forests for security reasons and also farmers who are near to them seek information frequently and those from a distance do not."*

A DA also supported this claim saying "*we get farmers who are near on a day-to-day bases and visit in two or three weeks' time those who are far*".

### 4.3.3. Digital Literacy

For the AEIS users, what is equally important to that of delivering digital content is adapting it to the literacy levels of the stakeholders. The stakeholders in the AEIS are composed of actors characterized by a mix of high- and low-level functional literacy, information literacy, and use of digital devices. For instance, key informants indicated their doubts on the capabilities of farmers to assimilate the information received to the context of their problems unless they are assisted how to put the information received into practice. In support of this, an informant, a project director at Bahardar University, working on information and knowledge provision to the Development Agents indicated the challenges they observed while using digital content as follows:

*Development agents have limitations in searching and using information and knowledge presented on digital devices and electronic sources.*

This excerpt shows the existing gaps among the stakeholders where those at the lower structure have issues of digital literacy to successfully apply the IS and use of digital devices.

## 5. DISCUSSION

The findings show that the current practice of the AEIS system is enabled with certain levels of resilience attributes with various information communication mechanisms. There are also significant challenges. Table 3 presents a summary of the mechanisms, challenges related to the mechanisms and the key transformations required to put a resilient IS for AEIS in place.

| Resilience attributes | Current mechanism | Challenge | Key Transformations |
|---|---|---|---|
| Robustness | Use of stakeholders and the social structures; local knowledge available within members; the ATAs' IVR-based IS | Lack of standard to apply local knowledge; too general information from the ATAs' IS | Consider empowerment of stakeholders at different levels to better serve their community |

| Self-organization | Stakeholders and members enabled to play roles in circulating contextually relevant information; discussions and supervision; continuous interaction | Role overlaps, limitation to support every farmer during seasons of intensive agricultural activity | Enabling actor networks and agency, and interaction of stakeholders |
|---|---|---|---|
| Learning | Use of social structures to facilitate horizontal and vertical information exchange; facilitated via direct observation, discussions, trainings, meetings, joint activities, supervision | Difficult for some smallholder farmers to apply information received and apply the mechanisms to every smallholder farmer | Enable agency of stakeholders (individual, proxy, and collective agencies) to enable interaction and participation |
| Redundancy | word-of-mouth public announcements, service encounters, publications, supervision, training, broadcast via radio and television, discussions, and joint activities | Highly scientific content through portals, reachability issue due to missed broadcasts | consider implementation at any level of the administrative structure (lower-level, middle-level, or top-level) |
| Rapidity | Mobile phone calls, Social and institutional structures; farmers associations; Board members | Problems to get access to some stakeholders | Consider stakeholders as information contributors and resource integrators |
| Scale | The multi-level farmers structures; institutional structures at various administrative levels | Increased complexity in communication as networks grow | Consider implementation at any level of the administrative structure |
| Diversity and Flexibility | Public announcements via word-of-mouth mechanisms, face-to-face discussions, service encounters, publication, and radios and television | Different functions and formats of information | Consider stakeholders literacies, and access to devices |
| Equality | Needs-based and voluntary service; the farmers social structures that involve every smallholder farmers | Lack of motivation to continuously seek information or to visit service encounters | Enable inclusive designs to incorporate every stakeholder into the service as beneficiaries |

**Table 3. Summary of the current mechanisms, challenges, and key transformations**

There is evidence of self-learning and development of capacity among the smallholder farmers and development agents to respond to their problems with no/little external support. The AEIS system is organized with identified roles to facilitate such learning. The social structure, namely the smallholder farmers' development groups, is purposefully designed to fill the void that would happen due to lack of access to expert services in the AEIS. This supports the findings of Haythornthwaite (1996) who reported actors in the social system as the sources for both information/knowledge and social support/influence. Such self-learning within the social actors allows inclusive and sustainable learning through sharing and adoption of diverse and collective social practices among themselves and with others (Kapuire et al., 2017; Kendall & Dearden, 2017). This implies that actors in the AEIS are not passive users who pull information for consumption or use information pushed but active participants of the functioning of the service. Enabling this in the IS designs for AEIS would provide the advantage suggested by Kendall & Dearden (2017), i.e., the shift from packaging and transmitting information towards facilitating communication among actors.

The self-organization resilience property in the AEIS community is a much needed property as agricultural practice is context specific. Hence a solution can be sustainable if it is tailored to individual fields or regions and can adapt to local agronomy practices and social conditions (Leeuwis, 2004). It is necessary to take the divergent concerns of actors in designing and building an effective system and to have a sense of shared ownership (Monk and Howard, 1998). This is possible when such interaction and supervision dominated service system operated by interconnected actors with defined roles is enabled. Such a structure can be easily adaptable to actors' networks and scalable to include new members and nodes. This is one of the markers of self-organization that can enable to find immediate solutions to problems. Actors' networks and complex interactions foster technical, social and institutional change in agricultural innovation (Klerkx et al., 2012). The effort to put equity resilience property ensures inclusive AEIS provision through creating equal access to agricultural extension information within the AEIS community.

Information provision through different mechanisms enables the diversity resilience property of IS (Sakurai & Kokuryo, 2014). However, IS which are not applied in the daily-life of actors are less effective to put resilient systems in place (Sakurai & Kokuryo, 2014). Mobile applications installed on devices which stakeholders are familiarized with can be considered useful in this regard. Application of such common infrastructure enabled by universality and ubiquity provides advantage of uniqueness and unison goals to enable resilient IS (Sakurai & Kokuryo, 2014). The given diversity improves the redundancy and scale resilience properties of information provision. Agriculture in its nature is considered as a co-evolutionary process involving interactive development of technology, practices, markets and institutions of growing network of actors (Klerkx et al., 2012) as a marker of scale. The stakeholder networks in the AEIS can be created and applied at different levels of administrative structures. This enables implementing redundant systems at different administrative structures (Kebele, Woreda, Region or National level) that could be accessible to other locations during times of shock. Though the above mechanisms are valuable in offering some properties of resilience there are challenges hindering to put a successful ISs for AEIS. For instance, information published on portals gets obsolete as it may not be updated as frequently as possible (Leeuwis, 2004).

The challenges and issues affect the equality in accessing the public service AEIS and prevent stakeholders from accessing services equally and to build inclusive AEIS systems. The problems related to the lack of design guidelines for the practitioners and the other issues reflected in the IS itself contributed to the lack of resilience features in the current ISs for AEIS. The ones, namely problems to incorporate content from stakeholders into existing systems, poor information transfer mechanisms, and lack/limited stock of services are critical to building resilient ISs for AEIS systems. The challenge of information transfer also include issues that made services inaccessible/unreachable to some sects of the community; the low commitment of the development agents and development group leaders, and other supporting institutions; and the travel distance that the service seeker or provider is expected to move or the geographical distance where the service centers are located. The social tie that a farmer has to key members in the social network and to experts in the organizational structure determines his/her access to AEIS. The low level of digital literacy explained by a combined low level of functional literacy, information literacy, and digital literacy complements the above challenges and the transformation towards a resilient IS for AEIS. These challenges attest the packaging and pushing of information from a center with an assumption of the receiver can change it to the desired outcomes do not hold true always. Because, information is expected to be processed or adapted by an actor in order to achieve the intended outcome (Wahid et al, 2017) and the mentioned challenges limit this capability. Transformations are required to advance the resilience of the IS for AEIS. Such transformation can be in the areas of designing the IS to enabling some capabilities and its implementation. The enabling aspect should consider empowerment of stakeholders at different levels, enabling actor-actor-networking and interaction, and enabling digital agency within the AEIS. Implementation of the design is required to consider some level of universality such as: possibility to implement the IS at any level of the AEIS structure, considering stakeholders as key content contributors and resource integrators, considering literacies levels of the target stakeholders and access to computing devices, and inclusion of the stakeholders to get access to basic public agricultural extension services.

## 6. Conclusion

This research applied the resilience framework to explore how the stakeholders of the Ethiopian AEIS system maintain resilience of information provision and the challenges and issues which could be considered as short and long term shocks) they encounter. Some of the transformation required to withstand and adapt the challenges are also identified. The resilience of IS for AEIS could be enhanced if they can facilitate interaction and communication among networked actors using devices stakeholders are familiarized with and taking into account their literacies and other characteristics. Therefore, building resilient IS for an AEIS system requires understanding and transforming foundational and enabling resilience attributes. Specifically, it is important to identify actors and actor-network along with their roles to enable ownership, interaction and communication within/across the social and institutional structures, and agency of actors.

The study contributes to the conversation on the application of the IS resilience framework in analyzing local information provision practices. By so doing, it adds to the empirical base of the ICT4D resilience framework literature and can be used to encourage and compare findings from other studies that follow the same line of enquiry. The results of this study can also contribute to kernel theories to develop IS design theory for AEIS. To practice, it highlights the key transformation areas to improve the effectiveness and impact of AEIS.

# REFERENCES AND CITATIONS